\documentclass[showpacs,showkeys,amsmath,amssymb,superscriptaddress,reprint, preprintnumbers,nofootinbib,notitlepage, groupdaddress,showabstract,aps,10pt,prd,notitlepage,showpacs,nofootinbib,superscriptaddress,compressed]{revtex4-1}

\pdfoutput=1 % if your are submitting a pdflatex (i.e. if you have
\usepackage[font=small]{caption}
\usepackage[utf8]{inputenc}
\usepackage[english]{babel}
\usepackage{bm}
\usepackage{graphicx}
\usepackage{dcolumn}
\usepackage{pbox}
\usepackage{epsfig}
\usepackage[dvipsnames]{xcolor}
\usepackage{slashed}
\usepackage{amssymb}
\usepackage{mathrsfs}
\usepackage{color}
\usepackage[font=small]{caption}
\usepackage[font=small]{subcaption}
\usepackage{url}
\definecolor{DarkBlue}{rgb}{0.7, 0.4, 1} 
\definecolor{Blue}{rgb}{0, 0.8, 0} 
\definecolor{MyLightBlue}{rgb}{0.5,0.7,1.9}
\definecolor{MyGreen}{rgb}{0.0,0.2, 0.0}
\definecolor{MyBrickRed}{rgb}{0, 0.5, 0.2}
\RequirePackage{hyperref}
\hypersetup{colorlinks, citecolor=Blue,linkcolor=DarkBlue, urlcolor=Green}
\usepackage[normalem]{ulem}

\newcommand{\bea}{\begin{eqnarray}}
\newcommand{\eea}{\end{eqnarray}}

\makeatletter

\renewcommand\@makecaption[2]{%
  \par
  \vskip\abovecaptionskip
  \begingroup
  
   \small\rmfamily
    \begingroup
     \samepage
     \flushing
     \let\footnote\@footnotemark@gobble
     \@make@capt@title{#1}{#2}\par
    \endgroup
  \endgroup
  \vskip\belowcaptionskip
}
\makeatother

\DeclareUnicodeCharacter{2212}{-}
\def\bea{\begin{eqnarray}}
\def\eea{\end{eqnarray}}
 \def\be{\begin{equation}}
\def\ee{\end{equation}}

\begin{document}
%\title{Effect of inelastic dark matter in hidden $U(1)$ extension of the Standard Model}
\title{Effect of inelastic scalar dark matter in hidden $U(1)$ scenario after the LZ nuclear recoil}
\author{Arindam Das}
\email{adas@particle.sci.hokudai.ac.jp}
\affiliation{Institute for Academic Innovation, Hokkaido University,
Sapporo, 060-0817, Japan}
\affiliation{Department of Physics, Hokkaido University, Sapporo 060-0810, Japan} 
\author{Takaaki Nomura}
\email{nomura@scu.edu.cn}
\affiliation{College of Physics, Sichuan University, Chengdu 610065, China}
%\date{\today}
%%%%%%%%%%%%%%%%%%%%%%%%%%%%%%%%%%%%%%%%%%%%%%%%%%%%%%%%%%%%%%%%%%%%%%%%%%%%%%%%%%%
\begin{abstract}
We study a hidden $U(1)_H$ extension of the Standard Model (SM) introducing two SM singlet complex scalar fields where one of them can be considered as a viable inelastic Dark Matter (DM) candidate due to its $U(1)_H$ charge being half of the charge of the other one which develops a vacuum expectation value. First we estimate a parameter region satisfying observed DM relic abundance through DM annihilation and further use these parameters to study inelastic DM-nucleus scattering to reproduce the recent high nuclear recoil energy event found by LZ collaboration. These interactions are mediated by a dark photon $(A^\prime)$ which is evolved after the spontaneous breaking of $U(1)_H$ symmetry and having a kinetic mixing with the SM photon. Hence we estimate bounds on the kinetic mixing as a function of $A^\prime$ mass and finally compare it with existing experimental bounds from different beam-dump and collider experiments. These parameters could be probed by lifetime and energy frontier experiments in future.
\end{abstract}
%%%%%%%%%%%%%%%%%%%%%%%%%%%%%%%%%%%%%%%%%%%%%%%%%%%%%%%%%%%%%%%%%%%%%%%%%%%%%%%%%%%%%
\maketitle

%%%%%%%%%%%%%%%%%%%%%%%%%%%%%%%%
\section{Introduction}
\label{sec:intro}
%%%%%%%%%%%%%%%%%%%%%%%%%%%%%%%%%
Except for the relic abundance \cite{Planck:2018vyg}, the nature and type of interaction about the most fundamental question of Dark Matter (DM) is unknown to us. Several proposals have been made regarding the nature and types of interactions of the DM \cite{Battaglieri:2017aum}, however, hitherto there is no clear signature. Recently  LUX-ZEPLIN (LZ) experiment
\cite{LZ:2026axp} announced the result of the extended search of events up to 270 keV nuclear recoil energy considering inelastic DM scattering with xenon nuclei from the data collection of 220 days in an exposure of 2.84 tonne-years. Compared to standard spin-independent (SI) scattering, high-energy events constitute a larger share of the expected recoil spectrum. The LZ collaboration then found one candidate event with nuclear recoil energy of 248$\pm$23(stat)$\pm$23(syst) keV with a global (local) significance of 2.6 (3.4)$\sigma$ where known expected background is low. 

To describe this scenario in the context of a simple phenomenological scenario we extend the SM with a hidden $U(1)_H$ gauge group \cite{Holdom:1985ag,Fayet:1990wx,Bjorken:1988as,Bauer:2018onh,Ilten:2018crw,Hambye:2018dpi,DallaValleGarcia:2025cwf} where all SM particles are neutral under this hidden $U(1)_H$ gauge group, while an extra SM singlet complex scalar field is charged under the hidden $U(1)_H$ gauge group. This scalar field acquires a vacuum expectation value (VEV) to originate the mass of dark gauge boson $A^\prime$. This dark gauge boson $A^\prime$ can interact with the SM particles through kinetic mixing between $U(1)_H$ and $U(1)_Y$ gauge field where $A'$ can be identified as dark photon. In addition to that, the SM Higgs and SM-singlet complex scalar fields are also mixed at the tree-level. In this framework we introduce another SM singlet complex scalar filed having half the charge of the previous SM singlet complex scalar field under the $U(1)_H$ gauge group. As a result, it can interact with the other BSM scalar through a triple scalar coupling and the $A^\prime$ gauge boson. Due to this charge assignment, the CP-odd and CP-even components of the second SM singlet complex scalar could have a mass difference. These CP-odd and even components can interact with the SM sector through the scalar mixing effect and $A^\prime$ mediation. The hidden $U(1)_H$ charge assignment of the second SM-singlet complex scalar ensures the stability of the CP odd and even components making them potential DM candidates and possibly classified as weekly interacting massive particles (WIMP) in nature \cite{Jungman:1995df,Arcadi:2017kky,Arcadi:2024ukq}. 
Remarkably the scalar DM candidates induce inelastic scattering via $A'$ mediation due to the mass difference and the type of interaction.

The recent finding of LZ \cite{LZ:2026axp} indicates the viability of an inelastic DM candidate which is theoretically well motivated \cite{Tucker-Smith:2001myb,Tucker-Smith:2004mxa,Okada:2019sbb,Li:2021rzt}. In the past experiments like DAMA/ LIBRA\cite{DAMA:2008jlt,DAMA:2010gpn} and XENON1t \cite{XENON:2020rca} generated some well motivated thoughts in the context of inelastic DM as done by the LZ result currently~\cite{Su:2026rwz,Freese:2026sga,Lou:2026idn,DiMauro:2026ldr,Yamashita:2026ump,Chattopadhyay:2026ryw,Du:2026guj,Dent:2026bji,Gu:2026vto,deLima:2026shq,Smirnov:2026aqk,Unwin:2026rdp,McCabe:2026crm,Yin:2026jnn,Jeesun:2026vzo,Rodd:2026tyn,Visinelli:2026kgt,Fan:2026kxx,Baer:2026fpy,Lee:2026wof,Borah:2026zwf,Wang:2026ytg,Yang:2026wpb,Kotlarski:2026pep,Liang:2026coz,DiMauro:2026dqp,Das:2026uyy,Alhazmi:2026efz,Ahmed:2026qjg,Du:2026lpa,Bandyopadhyay:2026gjw,Kannike:2026qyl,Bose:2026ndd,Elahi:2026vlm,Aghaie:2026vsu,Cheung:2026byg,Yuan:2026djt,Lee:2026xxh,Khan:2026nwp,Langhoff:2026ujr,Okada:2026eol, Bisal:2026khf,Chatterjee:2026scv,He:2026hqz,Fan:2026hzw,Chattaraj:2026fxn,Qi:2026vyp,Li:2026fci,Kumar:2026lgi,Frolovsky:2026tvq,Heikinheimo:2026kwp,Nguyen:2026lui,Lee:2026zbr,Okada:2026upm,Cabo-Almeida:2026uqw}. Conversely it has been pointed out that scenarios motivated by the LZ results could encounter astrophysical constraints \cite{Pospelov:2026ewn,Rodd:2026tyn}.

In the inelastic DM scenario there are two DM candidates with a mass splitting where the heavier state is considered as an excited state of the lighter state where their elastic interaction is either absent or assumed to be suppressed. Hence inelastic scattering of the DM state by the SM particles through the exchange of a mediator process dominates and thus it is converted to the excited state or the other way round. As a result the mass difference between these states play a crucial role in this process which is supposed to be constrained by 
direct detection (DD) experiment and residual DM annihilation from Cosmic Microwave Background (CMB). Additionally, when the mediator mass is small, i.e., $\mathcal{O}(< 1)$ GeV, light thermal DM scenario can be evolved satisfying 
Lee-Weinberg \cite{Lee:1977ua} bound. In the hidden $U(1)_H$ scenario, the CP-odd and CP-even states of the SM singlet complex scalar field could manifest this behavior while $A^\prime$ could be a mediator of the interaction.

In our paper, we consider a hidden $U(1)_H$ scenario where one of the two complex SM singlet scalars can be considered as a viable inelastic DM candidate due to its $U(1)_H$ charge being half of the $U(1)_H$ charge of the other complex SM singlet scalar; the case of inelastic fermion DM is also possible and it is discussed in ref.~\cite{Zhu:2026dag}. We study relic density through DM annihilation and compare this result with DM scattering with a nuclear target to study DD for inelastic DM satisfying current LZ nuclear recoil energy~\cite{LZ:2026axp}. The inelastic scattering process is mediated by $A^\prime$. In scalar DM case, we also have scalar interactions which affect relic density and DM-nucleus scattering, and we take into account these interactions discussing possible constraints. 
Then we discuss relevant parameters such as gauge kinetic mixing, gauge coupling, scalar couplings, scalar mixing and dark photon mass, which affect the DM physics like relic density and DM-nucleus scattering cross section. 
Hence using the LZ results we estimate constraints on the kinetic mixing parameter for spin independent (SI) DM-nucleus scattering fixing the DM mass and $U(1)_H$ gauge coupling which could be probed in future lifetime and energy frontiers.  

Our paper is organized as follows. We describe the model in Sec.~\ref{sec:mod}, DM physics in Sec.~\ref{sec:secIII}, results and discussions in Sec.~\ref{RD} and finally conclude the paper in Sec.~\ref{Conc}.
%%%%%%%%%%%%%%%%%%%%%%%%%%%%%%%%
\section{Model frameworks}
\label{sec:mod}
%%%%%%%%%%%%%%%%%%%%%%%%%%%%%%%%%

In this section, we summarize the framework of the model. In the model, hidden $U(1)_H$ gauge symmetry is introduced in addition to the SM gauge symmetry where all the SM fields are not charged under $U(1)_H$. We also add two complex scalar fields $S$ and $\varphi$ which have $U(1)_H$ charge $1$ and $2$ respectively.

The Lagrangian of the model is given by 
\begin{align}
\mathcal{L} = & \mathcal{L}_{\rm SM} - \frac{1}{4} X^{\mu \nu} X_{\mu \nu} - \frac{\epsilon'}{2} B_{\mu \nu} X^{\mu \nu} \nonumber \\
& + (D^\mu S)^*(D_\mu S) + (D^\mu \varphi)^*(D_\mu \varphi) -V,
\end{align}
where $\mathcal{L}_{\rm SM}$ is the SM Lagrangian without scalar potential, $B_{\mu \nu}$ and $X_{\mu \nu}$ are gauge field strengths for $U(1)_Y$ and $U(1)_H$ respectively, and $V$ represents scalar potential including the SM Higgs field $H$. The covariant derivative of scalar fields are written by
\begin{align}
& D_\mu S = (\partial_\mu - i g_H X_\mu) S, \\
& D_\mu \varphi = (\partial_\mu - 2 i g_H X_\mu) \varphi,
\end{align}
where $g_H$ is the gauge coupling associated with $U(1)_H$.
The scalar potential is given by 
\begin{align}
V = & - \mu_H^2 (H^\dagger H) - \mu^2_\varphi (\varphi^* \varphi) + M^2_S (S^* S) \nonumber \\
& - \mu (\varphi S^* S^* +c.c.) + \frac{\lambda_H}{2} (H^\dagger H)^2 + \frac{\lambda_\varphi}{2} (\varphi^* \varphi)^2 \nonumber \\
& + \frac{\lambda_S}{2} (S^* S)^2 + \lambda_{H\varphi} (H^\dagger H)(\varphi^* \varphi) \nonumber \\
& + \lambda_{H S} (H^\dagger H)(S^* S) +\lambda_{S \varphi} (S^* S)(\varphi^* \varphi).
\end{align}
The parameters in the potential is chosen to induce spontaneous electroweak and $U(1)_H$ symmetry breaking in which $H$ and $\varphi$ develop its VEVs; $\langle H \rangle = v/\sqrt{2}$ and $\langle \varphi \rangle = v_\varphi/\sqrt{2}$.
We write scalar fields as 
\begin{align}
& H = \begin{pmatrix} G^+ \\ \frac{1}{\sqrt{2}} (v + \tilde{h} + iG^0) \end{pmatrix}, \nonumber \\ 
& \varphi = \frac{1}{\sqrt{2}} (v_\varphi + \tilde{\phi} + i G'), \quad S = \frac{1}{\sqrt{2}} (s+ia), 
\end{align}
where $G^+$, $G^0$ and $G'$ are Nambu-Goldstone bosons absorbed by electroweak and hidden gauge bosons.

From our $U(1)_H$ charge assignment, there is remnant $Z_2$ symmetry where $s$ and $a$ are odd and other particles are even under it. Thus $s$ or $a$ can be DM candidate. 

\subsection{Massive neutral gauge bosons}
The kinetic terms are diagonalized by the transformation 
\begin{equation}
\begin{pmatrix} X_\mu \\ B_\mu \end{pmatrix} = 
\begin{pmatrix} r & 0 \\ - \epsilon' r & 1 \end{pmatrix}
\begin{pmatrix} \tilde{A}'_\mu \\ \tilde{B}_\mu \end{pmatrix},
\end{equation}
where $r=1/\sqrt{1-\epsilon'^2}$.
After spontaneous symmetry breaking, $Z$-$Z'$ mixing is induced such that 
\begin{align}
& \begin{pmatrix} \tilde{Z}_\mu \\ \tilde{A}'_\mu \end{pmatrix} = 
\begin{pmatrix} \cos \chi & \sin \chi \\ - \sin \chi & \cos \chi \end{pmatrix}
\begin{pmatrix} Z_\mu \\ A'_\mu \end{pmatrix}, \\ 
& \sin 2 \chi = \frac{2 \epsilon' \sin \theta_W m^2_{\rm SM}}{m_Z^2 - m^2_{A'}},
\end{align}
where $\theta_W$ is the Weinberg angle, $\tilde Z_\mu = \cos \theta_W W_\mu^3 - \sin \theta_W \tilde{B}_\mu$ with $W^3_\mu$ being the third component of $SU(2)_L$ gauge field and $m^2_{Z_{\rm SM}} = v^2 g^2/(4 \cos^2 \theta_W)$ with $U(2)_L$ gauge coupling $g$.
The $Z$-$Z'$ mixing angle $\chi$ is strongly suppressed as $|\chi| \lesssim 10^{-3}$ via electroweak precision test and $\epsilon' \ll 1$ is required. For small mixing, mass eigenvalues are approximated by 
\begin{equation}
m_Z \simeq m_{Z_{\rm SM}}, \quad m_{A'} \simeq 2 g_H v_\varphi.
\end{equation}
Also interaction among $A'$ and the SM particles are approximately given by
\begin{equation}
\mathcal{L}_{A'} = e \epsilon J^\mu_{\rm EM} A'_\mu,
\end{equation}
where $\epsilon \equiv \epsilon' \cos \theta_W$ and $J^\mu_{\rm EM}$ is the electromagnetic current. 
Thus $A'$ is identified as dark photon.

%%%%%%%%%%%%%%%%%%%%%%%%%%%%%%%%%%%%%%%%%%%%%%%%%%%%%%%%

\subsection{Scalar sector}
After spontaneous symmetry breaking, mass terms for $\phi$ and $\tilde{h}$ are given by
\begin{equation}
\frac{1}{2} \lambda_H v^2 h^2 + \lambda_{H\varphi} v v_\varphi h \phi + \frac{1}{2} \lambda_\varphi \phi^2.
\end{equation}
The corresponding mass matrix can be diagonalized by rotating $(h, \tilde{\phi})$ such that
\begin{align}
& \begin{pmatrix} \tilde{h} \\ \tilde{\phi} \end{pmatrix} =
\begin{pmatrix} \cos \alpha & -\sin \alpha \\ \sin \alpha & \cos \alpha \end{pmatrix} \begin{pmatrix} h  \\ \phi\end{pmatrix}, \label{eq:scalar1} \\
& \tan 2 \alpha = \frac{2 \lambda_{H \varphi}}{\lambda_H v^2 - \lambda_\varphi v_\varphi^2},
\end{align}
where we identify $h$ as the SM Higgs boson.
The mass eigenvalues are 
\begin{align}
& m_h^2 = \lambda_H v^2 \cos^2 \alpha + \lambda_\varphi v^2_\varphi \sin^2 \alpha + \lambda_{H \varphi} v v_\varphi \sin 2 \alpha, \\
& m_\phi^2 = \lambda_\varphi v_\varphi^2 \cos^2 \alpha + \lambda_H v^2 \sin^2 \alpha - \lambda_{H \varphi} v v_\varphi \sin 2 \alpha,
\end{align}
where $m_h \simeq 125$ GeV is the SM Higgs boson mass.

The $Z_2$ odd scalar boson masses are obtained as 
\begin{equation}
m^2_{s,a} = M^2_S + \frac{1}{4}(\lambda_{HS}v^2+\lambda_{S \varphi} v_\varphi^2) \mp \sqrt{2} \mu v_\varphi.
\end{equation}
The mass difference is then 
\begin{equation}
\delta m \equiv m_a - m_s = \frac{2 \sqrt{2} \mu v_\varphi}{m_a+m_s}.
\end{equation}

Interactions among scalar boson and $A'$ are written by~\cite{Nomura:2024bsz,Nomura:2024pwr}
\begin{align}
\mathcal{L} \supset & \frac{g_H}{2} A'_\mu (s \partial^\mu a - a \partial^\mu s) + \frac{g_H^2}{4} g_H^2 A'_\mu A'^\mu (s^2 +a^2), \nonumber \\
& + \frac{m_{A'}^2}{v_\varphi} A'^\mu A'_\mu (\phi \cos \alpha + h \sin \alpha) \nonumber \\
& + \chi \, m_{A'}^2\left(\frac{\cos \alpha}{v_\varphi^2}+ \frac{\sin \alpha}{v^2} \right)  \phi Z^\mu A'_\mu \nonumber \\
& + \chi \, m_{A'}^2\left(\frac{\sin \alpha}{v_\varphi^2}- \frac{\cos \alpha}{v^2} \right) h Z^\mu A'_\mu, 
\label{eq:gauge-int}
\end{align}
where we approximate $\sin \chi \simeq \chi$.
In addition, relevant interactions among scalar bosons are 
\begin{align}
\mathcal{L} \supset & \frac{\mu}{\sqrt{2}} \tilde{\phi} (s^2-a^2) - \frac{\lambda_{HS}}{4} (\tilde{h}^2 + 2 v \tilde{h})(s^2+a^2) \nonumber \\
& - \frac{\lambda_{S \varphi}}{4} (\tilde{\phi}^2 + 2 v_\varphi \tilde{\phi}) (s^2 + a^2) \nonumber \\
& + \sum_{\rm scalars}\lambda_{\varphi_i\varphi_j\varphi_k}\varphi_i\varphi_j\varphi_k, 
\end{align}
where $\tilde{h}$ and $\tilde \phi$ are given by Eq.~\eqref{eq:scalar1}.
The last term corresponds to the trilinear couplings for $Z_2$ even scalar bosons, and the relevant couplings are given by
\begin{align}
\lambda_{hhh} &= -\frac{m_h^2(v_\Phi \cos^3 \alpha  + v\sin^3 \alpha )}{2v v_\Phi}, \label{eq:hhh} \\
\lambda_{\phi hh} &= \frac{\sin 2\alpha (2 m_h^2 + m_\phi^2) (v_\Phi \cos \alpha  - v\sin \alpha )}{4 v v_\Phi}, \label{eq:phihh} \\
\lambda_{\phi \phi h} &=- \frac{\sin 2\alpha (m_h^2 + 2 m_\phi^2) (v \cos \alpha  +v_\Phi \sin \alpha )}{4 v v_\Phi} \label{eq:phiphih}.
\end{align}

%%%%%%%%%%%%%%%%%%%%%%%%%%%%%%%%%%%%%%%%%%%%%%%%%%%%%%%%%%%%%%%%%%%%%%%%%%%%%%%%%%%%%%%
\section{Dark matter physics}
\label{sec:secIII}
%%%%%%%%%%%%%%%%%%%%%%%%%%%%%%%%%%%%%%%%%%%%%%%%%%%%%%%%%%%%%%%%%%%%%%%%%%%

 %%%%%%%%%%%%%%%%%%%
\begin{figure}[tb]
\begin{center}
\includegraphics[width=62.0mm]{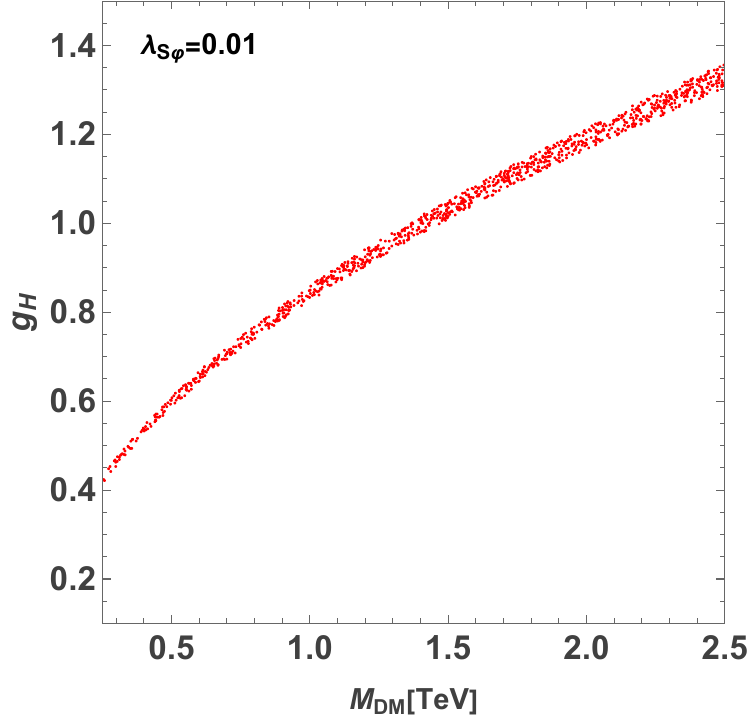} \\
\includegraphics[width=62.0mm]{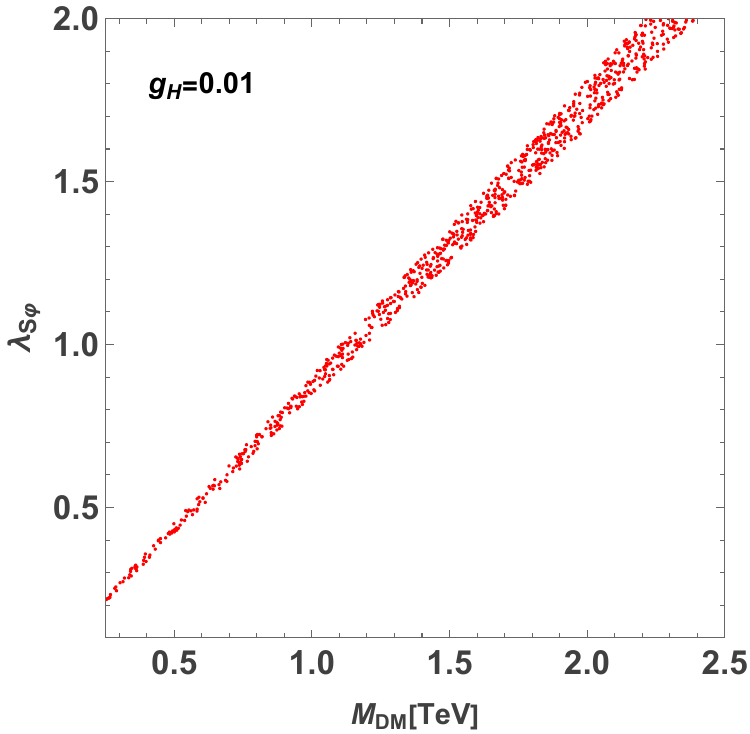} 
%%%
\caption{The parameter region that can accommodate observed relic density via the gauge interaction (top) and the scalar interactions (bottom).}
  \label{fig:relic}
\end{center}\end{figure}
%%%%%%%%%%%%%%%%%%%   

In this section, we discuss phenomenology regarding scalar DM in the model.

\subsection{Relic density}
Here we consider relic density of the DM.
The DM candidates can dominantly annihilate as $ss(aa) \to A' A'$ and $ss(aa) \to \phi \phi$ via gauge interaction and scalar interaction.
For $A'A'$ final state, DM annihilation cross section is approximately given by $\sigma v \sim g_H^4 /(16 \pi m^2_{\rm DM})$. On the ohter hand, for $\phi \phi$ final state, we obtain annihilation cross section $\sigma v \sim \lambda^2_{S \varphi}/(64 \pi m_{DM}^2)$ where we ignored $\mu$ coupling since we choose it to be small.
We estimate the relic density of DM adopting {\it micrOMEGAs7} \cite{Alguero:2023zol,Belanger:2026asz} implementing relevant parameters and interactions.

In Fig.~\ref{fig:relic}, we show the parameter region that can realize relevant relic density $0.11 < \Omega h^2 < 0.13$ for cases where $ss/aa \to A' A' [\phi \phi]$ mode is dominant in the left[right] panel. 
We find that $\mathcal{O}(1)$ coupling can accommodate observed relic density when DM is $\mathcal{O}(1)$ TeV.
Note that the relic density is insensitive to the mass of $A'$ as long as it is lighter than DM mass.

\subsection{Direct detection}

In the model, inelastic scattering occurs via gauge interaction Eq.~\eqref{eq:gauge-int} where dark photon is the mediator between DM and target nucleus $A$. 
The LZ event can be explained by the endothermic process $s A \to a A$ with $m_s < m_a$ where the minimal incoming speed of DM is required to be
\begin{equation}
v_{\rm min} = \sqrt{\frac{m_A E_R}{2 \mu^2_{sA}}} + \frac{\delta m}{\sqrt{2 m_A E_R}},
\end{equation}
where $m_A$ is the mass of target nucleus, $E_R$ is the recoil energy and $\mu_{sA} = m_s m_A/(m_s + m_A)$.
For LZ event, $E_R \simeq 248$ keV and $^{131}$Xe target, the minimal velocity reaches the end tail of the Galactic DM velocity distribution, explaining the absence of event with lower recoil energy. 
The LZ analysis~\cite{LZ:2026axp} shows a benchmark scenario in which the DM mass is TeV scale and DM scatters with target nucleus via spin-independent contact operator among DM and nucleon. Then mass splitting $\delta m = 350$ keV with 1 TeV DM can fit the data well. Using the benchmark scenario, we estimate the required spin-independent DM-nucleon scattering cross section, $\sigma_{SI}^{A'}$, in our case such that
\begin{equation}
2 \times 10^{-40} \ {\rm cm^2} < \sigma_{SI}^{A'} < 3 \times 10^{-39} \ {\rm cm^2} ,
\end{equation}
where we adopt $A^2/Z^2$ factor for LZ cross section since dark photon only couples to proton.

In the mode, we calculate the spin-independent cross section via dark photon interaction, and obtain
\begin{equation}
\sigma_{SI}^{A'} \simeq \frac{g_H^2 e^2 \epsilon^2}{2 \pi m^4_{A'}} \mu^2_{sN}, \label{eq:SI-cross-section}
\end{equation}
where $\mu_{sN} = m_s m_N/(m_s +m_N)$ with $m_N$ being nucleon mass. 

In addition to $A'$ mediation, we have a scalar mediating process that induces DM-nucleon elastic scattering. The cross section is estimated as
\begin{align}
\sigma_{SI}^{\rm scalar} \simeq & \frac{\mu^2_{sN}}{2 \pi} \frac{1}{m_s^2} \left( \frac{f_N m_N}{v} \right)^2 \nonumber \\
& \times\left( \frac{\lambda_{HS} v}{m_h^2} \cos \alpha - \frac{\lambda_{S \varphi} v_\varphi}{m_\phi^2} \sin \alpha \right)^2, \label{eq:elastic-CS}
\end{align}
where $f_N \sim 0.3$ is the Higgs coupling to the nucleon.
Note also that we ignored diagram with $\mu$ coupling since it is required to be small to obtain the small mass difference.
The cross section should be smaller than the constraints from the DM direct detection experiments.

 %%%%%%%%%%%%%%%%%%%
\begin{figure}[tb]
\begin{center}
\includegraphics[width=70.0mm]{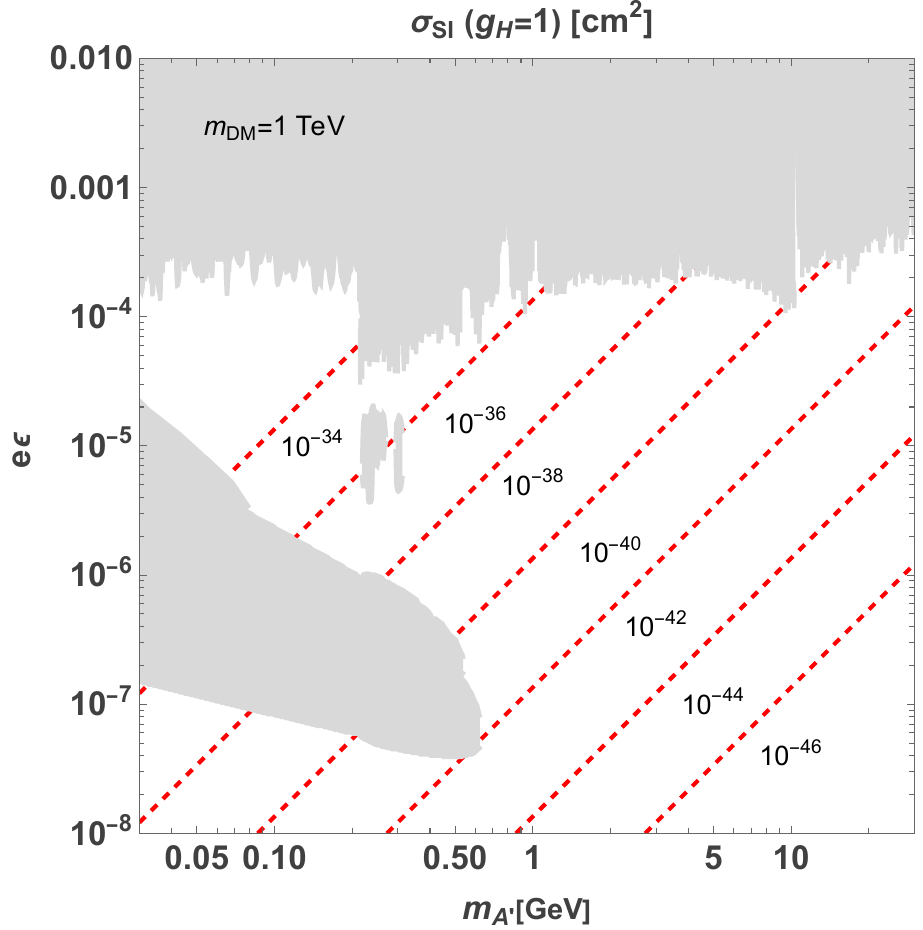} 
%%%
\caption{The contour shows spin-independent DM-nucleon scattering cross sections on $\{m_{A'}, e \epsilon \}$ plane. The gray shaded region is excluded by the dark photon searches.}
  \label{fig:DD1}
\end{center}\end{figure}
%%%%%%%%%%%%%%%%%%%   
%
 %%%%%%%%%%%%%%%%%%%
\begin{figure}[h!]
\begin{center}
\includegraphics[width=70.0mm]{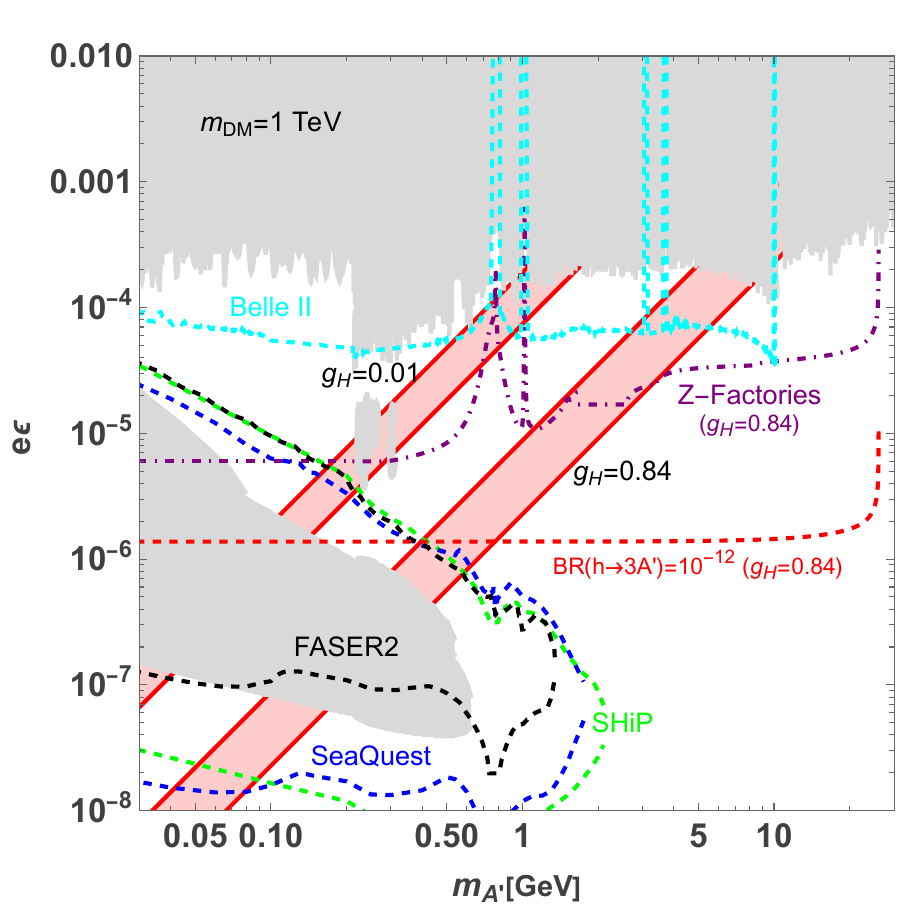} 
%%%
\caption{The red colored region indicate the parameter space that can explain the LZ event in a benchmark scenario $m_s = 1$ TeV and $\delta m = 350$ keV for $g_H =0.84$ and $g_H =0.01$. The sensitivity reaches expected at the future experiments are shown by the black (FASER2)~\cite{Feng:2017uoz,FASER:2018eoc}, green (ShiP)~\cite{SHiP:2015vad,SHiP:2020vbd}, blue (SeaQuest)~\cite{Berlin:2018pwi} and cyan (Belle II)~\cite{Belle-II:2018jsg} dashed curves. In addition, purple dot-dashed curve indicates the sensitivity reach by $h \to 3A'$ process at Z-factories with integrated luminosity of $10^5$ fb$^{-1}$ and $g_H = 0.84$, and red dashed curve indicates $BR(h \to 3Z') = 10^{-12}$~\cite{Nomura:2024bsz}.}
  \label{fig:DD2}
\end{center}\end{figure}
%%%%%%%%%%%%%%%%%%%   

%%%%%%%%%%%%%%%%%%%%%%%%%%%%%%%%%%%%%%%%%%%%%%%%%%%%%%%%%%%%%%%%
\section{Results and Discussions}
\label{RD}
%%%%%%%%%%%%%%%%%%%%%%%%%%%%%%%%%%%%%%%%%%%%%%%%%%%%%%%%%%%%%%%%

Here we provide our main results and discussions. 
Fig.~\ref{fig:DD1} shows the cross section Eq.~\eqref{eq:SI-cross-section} (for $g_H =1$) on $\{m_{A'}, e \epsilon \}$ plane. 
In the plot, the gray region is excluded by constraints from various dark photon search experiments such as beam dump experiments~\cite{Riordan:1987aw,Blumlein:1991xh,NA482:2015wmo,NA64:2019auh,NA62:2023qyn}, BarBar~\cite{BaBar:2014zli,BaBar:2016sci}, LHCb~\cite{LHCb:2017trq,LHCb:2019vmc} and FASER~\cite{Petersen:2023hgm} where the region is produced in use of {\it Darkcast} package~\cite{Ilten:2018crw}.
We find that sizable cross section can be realized for light $A'$ region avoiding the experimental constraints.

 %%%%%%%%%%%%%%%%%%%
\begin{figure}[t]
\begin{center}
\includegraphics[width=70.0mm]{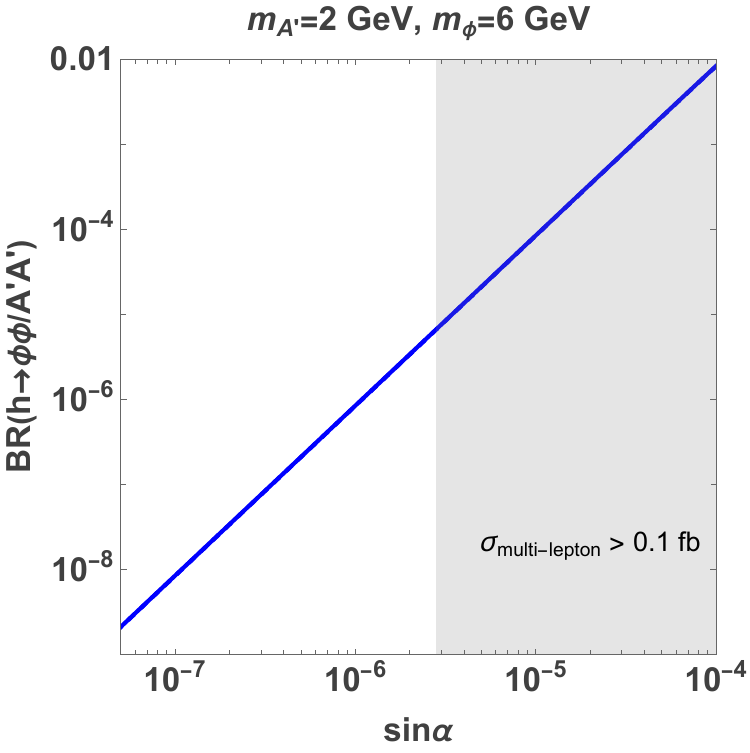} 
%%%
\caption{Branching ratio of $h \to \phi \phi/A'A'$ mode with $m_{A'} = 2$ GeV and $m_\phi =$ 6 GeV where $BR(h \to A'A' ) \simeq BR(h \to \phi \phi)$. Gray region is disfavored by 4 lepton search at the LHC with $\sqrt{s}=13$ TeV~\cite{ATLAS:2021wob}. }
  \label{fig:hDecay}
\end{center}\end{figure}
%%%%%%%%%%%%%%%%%%%   

Fig.~\ref{fig:DD2} expresses the region that can explain the LZ event for the benchmark scenario of $m_s = 1$ TeV and $\delta m = 350$ keV for different gauge coupling; relic density is explained by gauge interaction for $g_H =0.84$ and scalar interactions account for relic density for $g_H =0.01$.
The sensitivity reaches expected at the future experiments are shown by the black (FASER2)~\cite{Feng:2017uoz,FASER:2018eoc}, green (ShiP)~\cite{SHiP:2015vad,SHiP:2020vbd}, blue (SeaQuest)~\cite{Berlin:2018pwi} and cyan (Belle II)~\cite{Belle-II:2018jsg} dashed curves. In addition, purple dot-dashed curve indicates the sensitivity reach by $h \to 3A'$ process at Z-factories with integrated luminosity of $10^5$ fb$^{-1}$ and $g_H = 0.84$, and red dashed curve indicates $BR(h \to 3Z') = 10^{-12}$~\cite{Nomura:2024bsz}.
We find that relatively light dark photon $m_{A'} \lesssim 10$ GeV is preferred to explain the LZ event while avoiding experimental constraints.

Here we consider the constraint from elastic DM-nucleon scattering cross section Eq.~\eqref{eq:elastic-CS}. 
For $m_s = 1$ TeV, the cross section should be less than $\sim 3 \times 10^{-47} \ {\rm cm^2}$~\cite{LZ:2024zvo}. We obtain the constraint of $\lambda_{HS} \lesssim 0.04$ when the SM Higgs mediation is dominant choosing $\lambda_{S \varphi} \sin \alpha \to 0$. On the other hand, if $\phi$ mediation is dominant, we find the constraint 
\begin{equation}
\frac{v_\varphi^2}{\rm GeV^2} \frac{\rm GeV^4}{m^4_\phi} \lambda_{S\varphi}^2 \sin^2 \alpha \lesssim 3.7 \times 10^{-7}.
\end{equation}
We thus require $\lambda_{S\varphi}$ and/or $\sin \alpha$ to be small enough. 
It is also possible to realize small cross section via destructive interference between $h$ and $\phi$ mediating amplitude, but some fine tuning is required for the cancellation. 

An possible issue in the scenario is too long life-time of the next lightest dark particle, $a$ in our case, as discussed in the fermion DM case in ref.~\cite{Zhu:2026dag} since mass splitting is so small. It can be safe due to $a \to s + 3\gamma$ while the decay ratio is sensitive to mass difference $\delta m$. We can also consider adding light particle such as axion like particle (ALP) to make $a$ decay sufficiently fast.  

We also comment on the constraints from high energy solar neutrino when the DMs are captured by Sun~\cite{Pospelov:2026ewn,Nguyen:2026lui,Bose:2026ndd}. In our scenario, DM annihilate into $A'A'$ or $\phi \phi$ where $\phi$ also decays into $A'A'$. Since $A'$-neutrino coupling is suppressed the solar neutrino constraint would be relaxed in this case. Indeed, neutrino would be produced from decay chains via $A' \to \tau^+ \tau^-/\mu^+\mu^-/{\rm hadrons}$ depending on dark photon mass, and further analysis is required which is beyond the scope of this work.

Finally, we discuss relevant phenomenology associated with the scenario. In explaining the LZ event, relatively light $A'$ is required which can be tested in dark photon search experiments. In addition, extra scalar $\phi$ would be light since the scale of both $m_{A'}$ and $m_{\phi}$ are related to $v_\varphi$. In that case, we expect interesting events at collider experiments such as future Higgs/Z factories~\cite{FCC:2018byv, FCC:2025lpp, CEPCStudyGroup:2023quu, Ai:2025cpj, CEPCStudyGroup:2025kmw}. The Higgs factories can search process like $h \to A'A'$ and $h \to \phi \phi \to A' A'A' A'$ which induce multi-lepton final states.
The decay widths of these modes are summarized in the Appendx~\ref{sec:decay}.
In Fig~\ref{fig:hDecay}, we show branching ratio for $h \to A'A'$ and $h \to \phi \phi$ decay mode as a function of $\sin \alpha$ with benchmark values of $m_A' = 2$ GeV, $m_\phi = 6$ GeV and $g_H =0.84$; at the benchmark point $BR(h \to A'A') \simeq BR(h \to \phi \phi)$. The gray region is disfavored by 4 lepton serache at the LHC with $\sqrt{s} = 13$ TeV~\cite{ATLAS:2021wob}. 
We expect further parameter region can be explored in future Higgs factories.
In the Z factories, we also expect events such as $Z \to A' \phi \to A'A'A'$ where the decay width is given in the Appendix~\ref{sec:decay}.
We can test some region of our preferred parameter region in the future experiments where we show the expected sensitivity in Fig.~\ref{fig:DD2}. A detailed discussion regarding such signals from Higgs/Z decays can be referred to refs.~\cite{Li:2025luf,Nomura:2024bsz,Nomura:2024pwr,A:2024shl,Das:2025rlt}. We thus expect that some parameter regions to explain the LZ events are good target in future Higgs/Z factories.

%%%%%%%%%%%%%%%%%%%%%%%%%%%%%%%%%%%%%%%%%%%%%%%%%%%%%%%%%%%%%%%%
\section{Conclusions}
\label{Conc}
%%%%%%%%%%%%%%%%%%%%%%%%%%%%%%%%%%%%%%%
We study a hidden $U(1)_H$ extension of the SM with two SM singlet complex scalars $(S, \varphi)$ where one of the fields has half of the $U(1)_H$ charge of the other. The CP-even $(s)$ and odd $(a)$ multiplets of $S$ attain a mass splitting $(\delta m)$ after the spontaneous symmetry breaking. Due to this charge assignment the $s,$$a$ interact with the scalar sector and dark photon $(A^\prime)$, evolved after the spontaneous breaking of $U(1)_H$ gauge symmetry, of the model- making $s$ and $a$ potential inelastic DM candidate. Reproducing the observed DM relic abundance from DM annihilation and recently announced LZ nuclear recoil energy using inelastic DM- nucleus scattering we obtain $\delta m=350$ keV where DM mass is 1 TeV. 
Finally we estimate bounds on the gauge kinetic mixing parameter with respect to $m_{A^\prime}$ for DM mass at 1 TeV. We compare our results with the existing bounds on the gauge kinetic mixing from different beam dump and high energy collider experiments. Projected parameters of gauge kinetic mixing in this study could be probed by dark photon experiments in lifetime and energy frontiers involving FASER2, Large Hadron Collider at 3 ab$^{-1}$ luminosity, Forward search experiment at the 100 TeV hadron collider, beam dump facilities at the lepton colliders, etc. in future.

\begin{appendix}

\section{Decay widths of exotic Higgs and Z decay modes} \label{sec:decay}

Here we summarize some decay widths for exotic Higgs and Z decays~\cite{Nomura:2024bsz,Nomura:2024pwr}. 
The $h \to A'A'$ and $h\to \phi \phi$ modes can be calculated via interactions Eq.~\eqref{eq:gauge-int} and \eqref{eq:phiphih}.
We then obtain
\begin{align}
\begin{split}
& \Gamma (h \to A' A') = \frac{m_h^3 \alpha^2}{32 \pi v_\varphi^2} + \mathcal{O}(\alpha^4), \\
& \Gamma(h \to \phi \phi) = \frac{m_h^3\alpha^2}{32 \pi v_\varphi^2} \left(1 + \frac{2 m_\phi^2}{m_h^2} \right)^2 \left(1 + 2\alpha  \frac{v_\varphi}{v}\right) + \mathcal{O}(\alpha^4).
\end{split}
\label{eq:width-limit}
\end{align}
For $Z \to A' \phi$, the decay width can be calculated using the interaction Eq.~\eqref{eq:gauge-int}.
We can write it as
\begin{align}
&\Gamma(Z \to A'\phi) = 
 \frac{m_Z^3}{48\pi }\left(\sin\alpha + \frac{v}{v_\Phi^{}}\cos\alpha \right)^2 x_{Z^\prime}^{}\chi^2\nonumber \\
&~~~ \times \lambda^{1/2}(x_{Z'},x_\phi)\left[(1 + x_{Z'} - x_\phi)^2 + 8x_{A'}\right],
\end{align}
where $x_{A'} = m_{A'}^2/m_Z^2$, $x_{\phi} = m_{\phi}^2/m_Z^2$ and $\lambda(x,y) = 1+x^2+y^2-2xy-2x-2y$.

\end{appendix}

%%%%%%%%%%%%%%%%%%%%%%%%%
\vspace{-0.0098 in}
\bibliographystyle{utphys}
\bibliography{bibliography}
%%%%%%%%%%%%%%%%%%%%%%%%%%%
\end{document}